\documentclass[aps,pra,twocolumn]{revtex4-2}
\usepackage{amsmath,amssymb,bm}
\usepackage{graphicx}
\usepackage{hyperref}
\usepackage{physics}

\begin{document}

\title{Two mechanisms of backward optical forces on Rayleigh particles in structured paraxial light}

\author{Tomasz Rado\.zycki}
\affiliation{Institute of Physical Sciences, Cardinal Stefan Wyszynski
University in Warsaw, W\'oycickiego 1/3, 01-938 Warsaw, Poland}

\date{\today}
\begin{abstract}
A theoretical and numerical study of optical forces acting on a Rayleigh particle in a paraxial Gaussian light beam exhibiting regions of optical backflow is presented. Within the dipole approximation, the total optical force is decomposed into gradient, scattering, and spin-curl terms. Vector fields satisfying the exact paraxial Maxwell equations are employed to describe the structured light configuration responsible for two distinct mechanisms leading to backward optical forces. The first originates from the local reversal of the Poynting vector, which induces a negative longitudinal momentum flux, while the second arises from the spin-dependent component of the force associated with the spatial variation of the optical spin density. Analytical expressions and numerical simulations confirm that both mechanisms can produce backward motion of a Rayleigh particle under appropriate beam conditions. These results provide a unified physical picture of backward-directed optical forces in Gaussian beams and open possibilities for particle manipulation in structured light fields.
\end{abstract}

\maketitle

\section{Introduction}

Using light to manipulate tiny particles has become a key technique in modern optics and nanotechnology \cite{ash,chu,miller,pad,gaon,kolbow}. For very small (Rayleigh) particles, optical forces are usually described by two main components: the gradient force, which pulls particles with positive polarizability toward regions of higher intensity (or repels particles with negative polarizability), and the scattering force, which pushes particles along the light propagation direction \cite{neuman,jones}. In many cases, these two forces are sufficient to describe particle motion.

In structured light fields, however, the situation becomes more complex. Some regions exhibit optical backflow, where the longitudinal component of the Poynting vector is negative, indicating a local reversal of energy flow \cite{saari,ghg,ibba,elie,daniel,trback,trmax}. In these regions, nonconservative forces can act against the beam’s propagation, forming the basis for optical pulling or tractor beams \cite{nov,alba,shlee,chen,sukhov,yev}. Backward motion can also arise from variations in the transverse spin density, where the curl of the spin induces local off-axis contributions, even though the overall $z$-directed spin flux remains zero. Such effects illustrate the subtle interplay between field topology and mechanical action: even in a globally forward-propagating beam, structured phase and polarization distributions can locally redirect momentum flow. This phenomenon highlights that optical forces are governed not only by intensity gradients but also by the vectorial structure of light.

These observations lead to a natural decomposition of the total optical force into a conservative gradient component and nonconservative scattering and spin-curl components. Such a decomposition clarifies how structured light interacts with particles, explaining both conventional forward motion and the unusual backward-directed trajectories observed in backflow regions.

Experimental studies have confirmed that optical pulling forces can occur in diverse systems, including structured light fields and surface plasmon polaritons \cite{ruf,kaj,ding,pet,gong,lee,wang}, enabling new applications in remote particle transport, optical sorting, and precise positioning of nanoparticles. Despite these advances, the fundamental mechanisms leading to backward-directed motion in paraxial fields remain only partially understood. In particular, the relative roles of scattering and spin-curl contributions have not been systematically examined within a unified theoretical framework. Addressing this gap requires combining analytical models with direct numerical simulations of particle trajectories.

In this paper, we present a theoretical and numerical study of optical forces on a Rayleigh particle in a special paraxial Gaussian beam exhibiting backflow regions. Section~\ref{teor} outlines the theoretical basis and the field structure responsible for backflow. Section~\ref{foran} derives the dimensionless equation of motion, decomposes the total force into gradient, scattering, and spin-curl terms, and illustrates these components providing a coherent physical picture of the mechanisms involved. Section~\ref{numtr} presents numerical simulations of particle trajectories, explicitly showing the backward motion due to the Poynting-vector backflow and spin–momentum coupling.

\section{Theoretical framework}\label{teor}

Within the dipole approximation, the optical force acting on a polarizable Rayleigh particle, in terms of  real instantaneous fields and real dipole moment $\bm{p}$, is given by~\cite{coi} 
\begin{equation}\label{forcein}
\bm{F}(t)=(\bm{p}\cdot\bm{\nabla})\bm{E}(t)+\dot{\bm{p}}\times\bm{B}(t)+\bm{p}\times\left[(\bm{v}\cdot\bm{\nabla})\bm{B}(t)\right].
\end{equation}
Using the identity $\bm{\nabla}\cdot\bm{B}=0$, the last term is often rewritten in the form of $\bm{v}\times(\bm{p}\cdot\bm{\nabla})\bm{B}$. Since the motion of Rayleigh particles in light beams is very slow (especially when compared to the speed of light) this term plays only marginal role and is, therefore, neglected in the following. 

Let us now consider the monochromatic beam in vacuum. Upon introducing time-independent complex fields $\bm{E}_c$ and $\bm{B}_c$ according to
\begin{equation}\label{erebre}
\bm{E}(t)=\Re\left(\bm{E}_ce^{-i\omega t}\right),\qquad \bm{B}(t)=\Re\left(\bm{B}_ce^{-i\omega t}\right),
\end{equation}
 as well as complex polarizability $\alpha_c=\alpha_R+i\alpha_I$, where the subscript $c$ stands for ``complex'', force (\ref{forcein}), averaged over field oscillations, can be given the form \cite{boulo}
\begin{eqnarray}\label{sius}
\expval{\bm{F}} &=& \frac{1}{2}\sum_j \Re\left[\alpha_c E_{cj} \bm{\nabla} E_{cj}^* \right]= \frac{1}{4}\,\alpha_R\bm{\nabla} \Big(\sum_j |E_{cj}|^2\Big)\nonumber \\
&&+\;\omega\mu_0\alpha_I\expval{\bm{S}}
-\frac{1}{4}\,\alpha_I\bm{\nabla}\times\Im \left(\bm{E}_c^*\times\bm{E}_c\right).
\end{eqnarray}
Here $\bm{p}_c = \alpha_c \bm{E}_c$ is the induced dipole moment, $\expval{\bm{S}}=\frac{1}{2\mu_0}\,\Re\left(\bm{E}_c\times \bm{B}_c^*\right)$ i.e., stands for the time-averaged Poynting vector,  and $\mu_0$ denotes the vacuum permeability. The averaging brackets 
$\expval{}$ will be omitted hereafter. 
Formula (\ref{sius}) represents the decomposition of the total force into three contributions --
gradient, scattering, and spin-curl forces: 
\begin{subequations}\label{rosi} 
\begin{align}
&\bm{F}_{\text{grad}} = \frac{1}{4}\,\alpha_R\bm{\nabla} \Big(\sum_j |E_{cj}|^2\Big), \label{rosigr}\\
&\bm{F}_{\text{scatt}} = \omega\mu_0\alpha_I\bm{S}, \label{rosis}\\
&\bm{F}_{\text{spin}} = -\frac{1}{4}\,\alpha_I\bm{\nabla}\times\Im \left(\bm{E}_c^*\times\bm{E}_c\right)\label{rosisp}.
\end{align}
\end{subequations}
$\bm{F}_{\text{grad}}$ is the fundamental force responsible for trapping particles in regions of high ($\alpha_R>0$) or low ($\alpha_R<0$) intensity, $\bm{F}_{\text{scatt}}$ corresponds to the radiation pressure, whereas $\bm{F}_{\text{spin}}$ arises from the spatial variation of the field’s spin angular momentum and can exert both a torque, causing the particle to rotate, and a translational force associated with spin–curl gradients. The forces still depend on time, not because of the temporal variation of the fields -- which has already been averaged out -- but due to the changing position of the particle as it moves within these fields.

The backward force may arise either from $\bm{F}_{\text{scatt}}$, when the Poynting vector exhibits backflow, or from $\bm{F}_{\text{spin}}$ when polarization inhomogeneity is dominant (the gradient force 
points toward local intensity maxima and may have backward components, which indicate trapping rather than the reverse motion). Typically, the former is by far larger than the latter. In this study, in order to examine the effects in question, we employ the example of a paraxial Gaussian beam described in \cite{trback}, where the Poynting vector reveals local areas of backward energy propagation.

To simplify the analysis, it is convenient to start by introducing dimensionless variables:
\begin{equation}\label{resca}
\bm{\xi}=k\bm{r},\qquad \tau=\omega t,
\end{equation}
where $k=\omega/c$. Furthermore, dimensionless electromagnetic fields can be defined by factoring out the dimensional constants  $E_0$ and $B_0$ (which can be chosen to be real) associated with the overall beam intensity:
\begin{equation}\label{polebe}
\bm{E}_c=E_0\bm{\mathcal{E}},\qquad \bm{B}_c=B_0\bm{\mathcal{B}}.
\end{equation}
Their exact values are inessential to the present analysis, since all three force components scale with their squares. 

Before specifying the explicit form of the fields $\bm{\mathcal{E}}$ and $\bm{\mathcal{B}}$, let us introduce a small dimensionless parameter characterizing the degree of beam focusing:
\begin{equation}\label{parep}
\varepsilon=\frac{1}{kw_0}=\frac{\lambda}{2\pi w_0}=\frac{w_0}{2z_R},
\end{equation}
where the so-called Rayleigh range is defined as $z_R=kw_0^2/2$. Then $kz_R=1/(2\varepsilon^2)$, establishing a direct relation between the Rayleigh range and the parameter $\varepsilon$. An increase in 
$\varepsilon$ thus implies a reduction in the Rayleigh range, corresponding to a stronger beam focusing.

According to \cite{trback}, the two potentials $V_\pm(\bm{\xi})$ are chosen in the form
\begin{subequations}\label{vv}
\begin{align}
&V_+(\bm{\xi})=\frac{i\varepsilon\xi_y}{(1+2 i\varepsilon^2 \xi_z)^2} e^{-\frac{\varepsilon^2  (\xi_x^2+\xi_y^2)}{1+2 i\varepsilon
   ^2 \xi_z}},\label{vvp}\\
&V_-(\bm{\xi})=\frac{-\varepsilon\xi_x}{(1+2 i\varepsilon^2 \xi_z)^2} e^{-\frac{\varepsilon^2  (\xi_x^2+\xi_y^2)}{1+2 i\varepsilon
   ^2 \xi_z}}.\label{vvm}
\end{align}
\end{subequations}
describing the slowly varying envelopes. They obey the standard scalar paraxial propagation condition, which in the dimensionless coordinates introduced earlier reads:
\begin{equation}\label{rps}
\left[\partial^2_{\xi_x}+\partial^2_{\xi_y}+2i\partial_{\xi_z}\right]V_\pm(\bm{\xi})=0.
\end{equation}
The obvious abbreviation $\partial_{\xi_x} \equiv \partial / \partial \xi_x$ is used throughout, and analogously for other variables. 

Condition (\ref{rps}) ensures that the electromagnetic fields, defined by the following differentiations, satisfy the paraxial Maxwell equations (i.e., the vectorial form) formulated in \cite{trmax} and referred to throughout this paper as the {\em exact paraxial equations}. The electric field components are given by
\begin{subequations}\label{parpee}
\begin{align}
&\mathcal{E}_{x}=e^{i\xi_z}\Big[\partial_{\xi_x}\Big(1-\delta^2\,\frac{i}{2}\,\partial_{\xi_z} \Big)V_+\nonumber\\
&\qquad-i\partial_{\xi_y}\Big(1+\delta^2\,\frac{i}{2}\,\partial_{\xi_z}\Big)V_-\Big],\label{parpexe}\\
&\mathcal{E}_{y}=e^{i\xi_z}\Big[i\partial_{\xi_x}\Big(1+\delta^2\,\frac{i}{2}\,\partial_{\xi_z} \Big)V_-\nonumber\\
&\qquad+\partial_{\xi_y}\Big(1-\delta^2\,\frac{i}{2}\,\partial_{\xi_z}\Big)V_+\Big],\label{parpeye}\\
&\mathcal{E}_{z}=2\delta\,e^{i\xi_z}\partial_{\xi_z}\widetilde{V}_+,\label{parpeze}
\end{align}
\end{subequations}
and the magnetic field components are expressed as
\begin{subequations}\label{parpbe}
\begin{align}
&\mathcal{B}_{x}=e^{i\xi_z}\Big[-i\partial_{\xi_x}\Big(1-\delta^2\,\frac{i}{2}\,\partial_{\xi_z} \Big)V_-\nonumber\\
&\qquad-\partial_{\xi_y}\Big(1+\delta^2\,\frac{i}{2}\,\partial_{\xi_z}\Big)V_+\Big],\label{parpbxe}\\
&\mathcal{B}_{y}=e^{i\xi_z}\Big[\partial_{\xi_x}\Big(1+\delta^2\,\frac{i}{2}\,\partial_{\xi_z} \Big)V_+\nonumber\\
&\qquad-i\partial_{\xi_y}\Big(1-\delta^2\,\frac{i}{2}\,\partial_{\xi_z}\Big)V_-\Big],\label{parpbye}\\
&\mathcal{B}_{z}=-2i\delta\, e^{i\xi_z}\partial_{\xi_z}\widetilde{V}_-.\label{parpbze}
\end{align}
\end{subequations}
The parameter $\delta$ formally equals unity, but it was introduced to facilitate tracing the effect of additional terms appearing in the refined paraxial approximation, in which the fields exactly satisfy the paraxial Maxwell equations. It thus acts as a switch that allows one to turn these terms on and off. In particular, setting $\delta=1$ is essential whenever physical effects originating from longitudinal field components and spin–orbit coupling are considered, as these contributions vanish identically for 
$\delta=0$ (see, for instance, forces in Fig. \ref{forces2d}). In contrast, the leading-order intensity profile and the conventional gradient force are already correctly captured in the limit.

It should be emphasized that the above field expressions represent refined paraxial beam models that are fully consistent with Maxwell’s equations within the paraxial framework, rather than exact nonparaxial solutions. From an experimental perspective, beams described by the exact paraxial equations can be realized using standard techniques of modern singular and vector beam optics. 
In practice, the required spatially structured amplitude and polarization fields can be generated by combining spatial light modulators (SLMs) with polarization control elements, such as wave plates or $q$-plates, which allow independent shaping of the scalar envelope and the local polarization state. 
Such approaches are routinely employed for the generation of higher-order Gaussian and Laguerre--Gaussian modes, vector beams, and beams with tailored phase and amplitude profiles, whose slowly varying envelopes satisfy the paraxial wave equation to high accuracy. 
The present fields, therefore, represent not an abstract mathematical construction, but a realistic extension of experimentally accessible paraxial beams in which higher-order vectorial corrections become relevant~\cite{mair,forbes,shen,forb}.

\begin{figure}[h!]
\centering
\includegraphics[width=0.48\textwidth,angle=0]{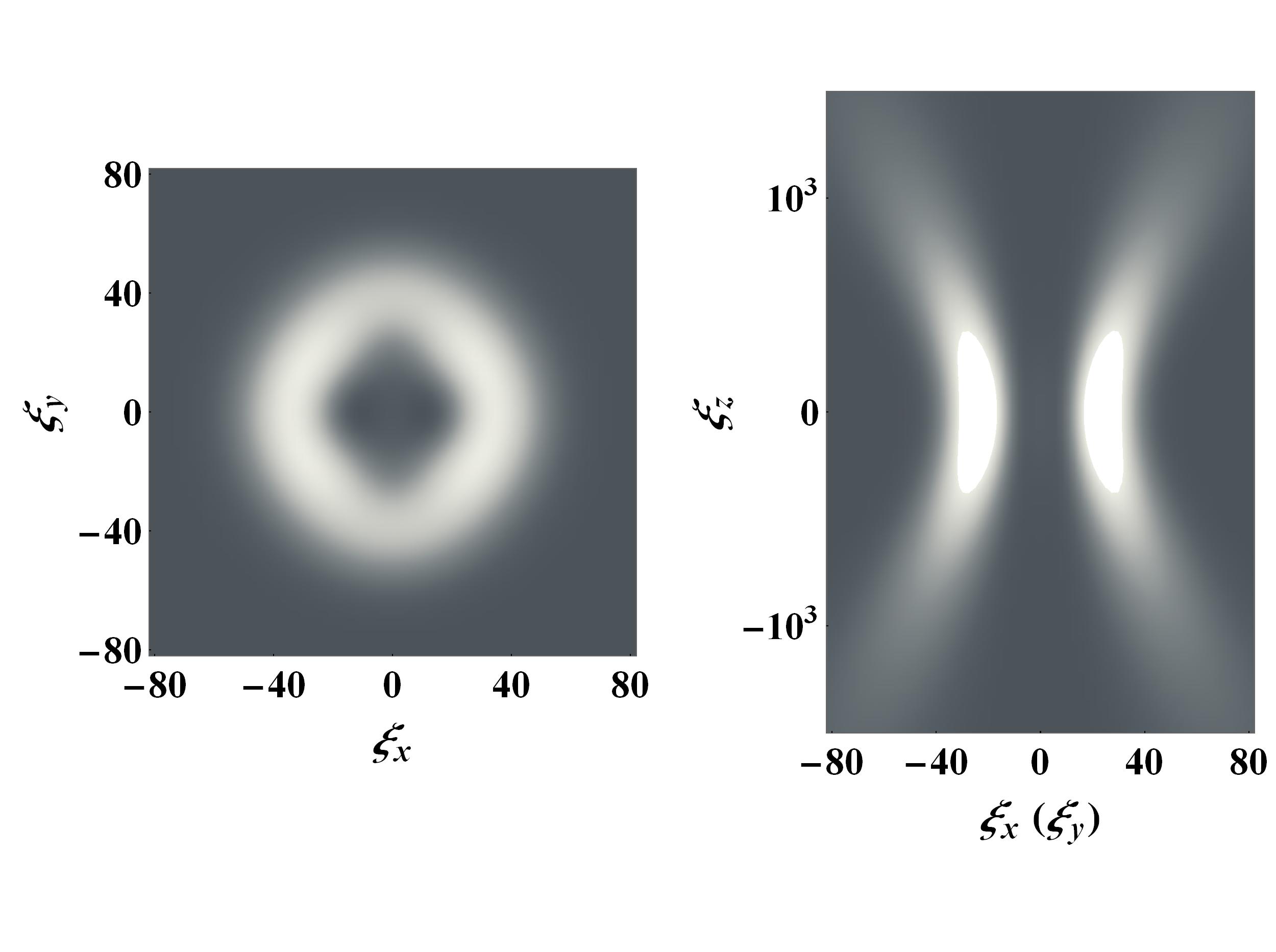}
\vspace{-35pt}
\caption{Wave intensity for the fields (\ref{parpee}) and (\ref{parpbe}) in the plane $\xi_z=0$ (left plot), and $\xi_x=\xi_y$ (right plot) for $\varepsilon=0.03$ and $\delta=1$. Brighter areas correspond to higher intensity.}
\label{inty}
\end{figure}

For these fields, the intensity profiles in the focal and axial planes are shown in Fig.~\ref{inty}. The distributions exhibit features characteristic of a first-order Gaussian beam (or Laguerre–Gaussian beam), although the observed intensity ring displays some asymmetry. 
It is not accidental and originates already at the level of the scalar potentials $V_\pm$, which are not axially symmetric due to their explicit dependence on $\xi_x$ and 
$\xi_y$. This intrinsic asymmetry is further enhanced by the vectorial (non-scalar) corrections included in the exact paraxial formulation. 

The bright regions of high intensity can serve as potential traps for particles with a positive real polarizability, as follows from the expression for the gradient force, whose direction points toward the steepest increase in the field intensity.

\begin{figure}[h!]
\centering
\includegraphics[width=0.45\textwidth,angle=0]{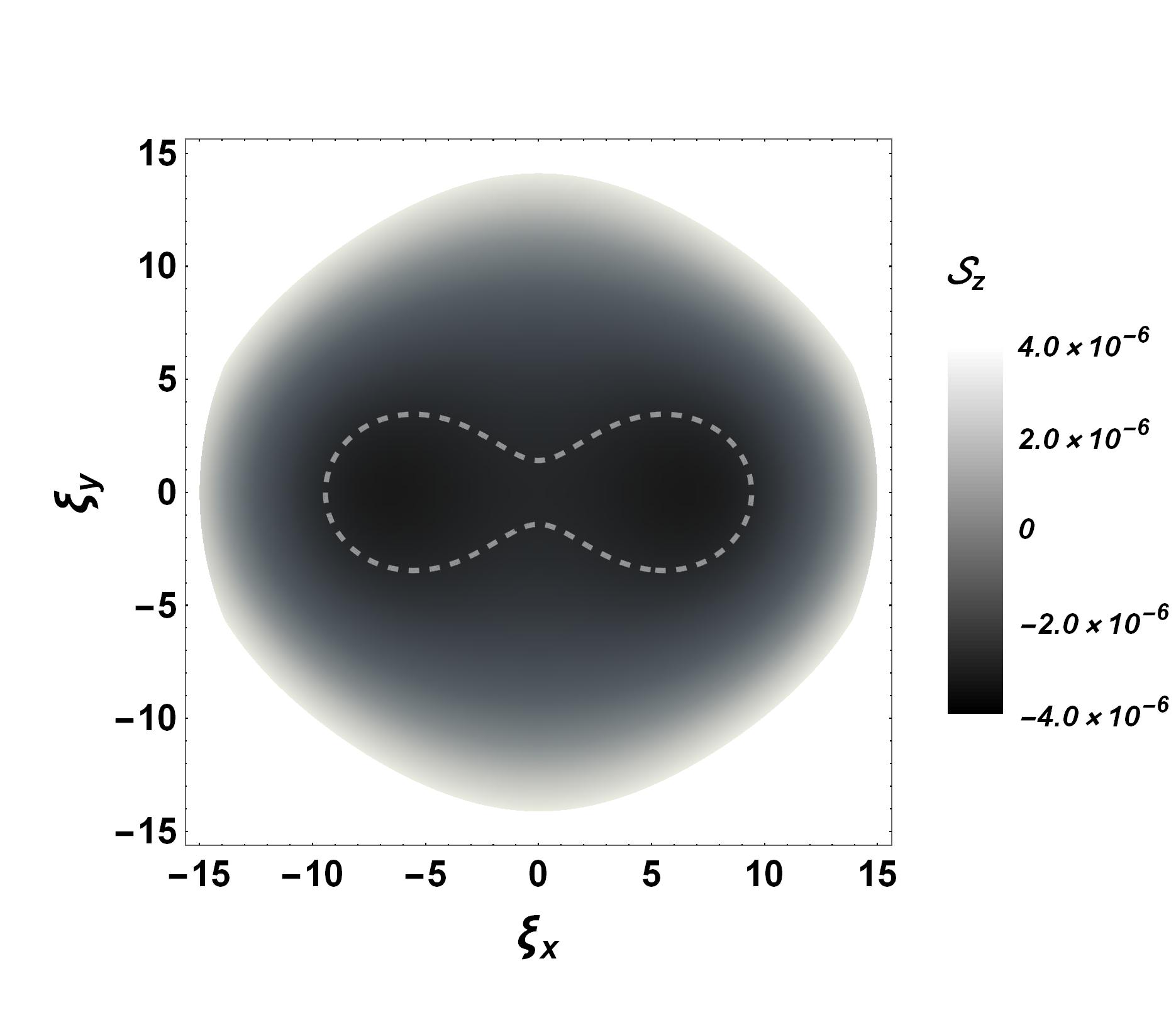}
\caption{The dimensionless $z$ component of the Poynting vector, as appearing in Eq.~(\ref{rosics}) in the plane $\xi_z=0$ for $\varepsilon=0.03$ and $\delta=1$. The dashed line marks the region of optical backflow.}
\label{backs}
\end{figure}

Of particular importance for our purposes is that the beam constructed in this manner also exhibits optical backflow in certain spatial regions \cite{trback}. In these regions, the local Poynting vector develops components directed opposite to the overall propagation of the beam, resulting in a local reversal of the energy flow. This distinctive feature makes the beam an excellent model for analyzing both conventional and anomalous contributions to the optical force, which are central to the present study. The distribution of the $z$ component of the Poynting vector [rescaled to its dimensionless form as in Eq.~(\ref{rosics})] is shown in Fig.~\ref{backs}.

The Rayleigh particle is modeled as a point dipole characterized by its mass $m$ and complex polarizability $\alpha_c$. This approximation is valid when the particle size is much smaller than the distance over which the optical field distribution undergoes significant variation. In the quasi-static (Rayleigh) limit, the electrostatic polarizability of a homogeneous, non-magnetic sphere of radius $R$ and refractive index $n$ is given by the Lorenz-Lorentz formula  
\begin{equation}\label{cmfor}
\alpha = 4\pi\epsilon_0 R^3 \frac{n^2-1}{n^2+2},
\end{equation}
which follows from solving Laplace’s equation for a small sphere in a uniform field, and setting the induced dipole moment to $\bm{p} = \alpha \bm{E}$.
The Lorenz-Lorentz formula is valid provided the particle is much smaller than the wavelength ($kR\ll 1$), homogeneous and isotropic, and multiple-scattering and near-field corrections can be neglected (quasi-static approximation).

To account for radiation damping (and, if relevant, intrinsic material absorption), the dynamic (radiation-corrected) polarizability can be used \cite{alba2}:
\begin{equation}\label{alfaco}
\alpha_c=\frac{\alpha}{1-\dfrac{i}{6\pi\epsilon_0}\,k^3\alpha}
        \equiv\frac{\alpha}{1-i\beta},
\qquad \beta\equiv\frac{1}{6\pi\epsilon_0}\,k^3\alpha.
\end{equation}
The imaginary term in the denominator arises from the self-reaction of the radiating dipole and enforces consistency with the optical theorem (it ensures that scattering losses are accounted for). 
Writing $\alpha_c=\alpha_R+i\alpha_I$ one obtains
\begin{equation}\label{alui}
\alpha_R=\frac{\alpha}{1+\beta^2},\qquad
\alpha_I=\frac{\alpha\beta}{1+\beta^2}=\alpha_R \beta.
\end{equation}

Physically, the real part $\alpha_R$ determines the conservative (gradient) response of the dipole and thus controls trapping: for $n>1$ we have $\alpha>0$, so $\alpha_R>0$ and the gradient force pulls the particle toward intensity maxima.

The imaginary part $\alpha_I$ encodes dissipative processes: it accounts both for radiative damping (scattering) and, if present, material absorption. 
For transparent dielectric particles, intrinsic absorption is negligible, and $\alpha_I$ is dominated by radiation reaction.

\section{Analysis of forces}\label{foran}

\begin{figure}[h!]
\centering
\includegraphics[width=0.4\textwidth,angle=0]{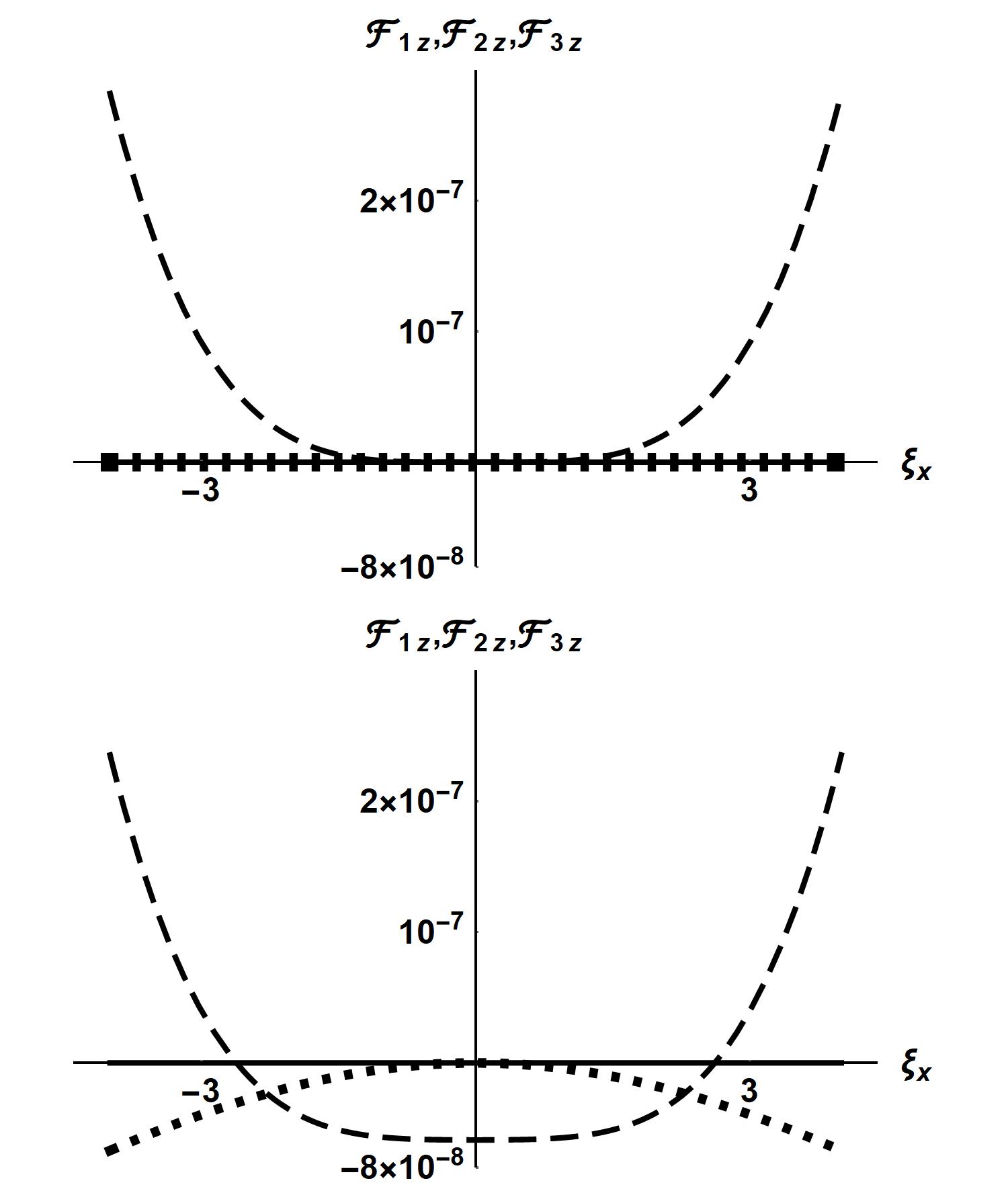}
\caption{Comparison of the axial components of $\mathcal{F}_{\text{grad},z}$ (solid line), $\mathcal{F}_{\text{scatt},z}$ (dashed line), and $\mathcal{F}_{\text{spin},z}$ (dotted line). The forces are plotted along $\xi_y = -\xi_x$ for $\xi_z =0$, with regions of dominance for each component clearly distinguishable. Parameter values are $\varepsilon = 0.03$, $\gamma = 1$, and $\beta = 0.05$. The lower panel shows the forces obtained from the full expressions~(\ref{parpee}) and~(\ref{parpbe}), whereas the upper panel corresponds to the case $\delta = 0$, where all backward-directed forces vanish.}
\label{forces2d}
\end{figure}

\begin{figure}[h!]
\centering
\includegraphics[width=0.485\textwidth,angle=0]{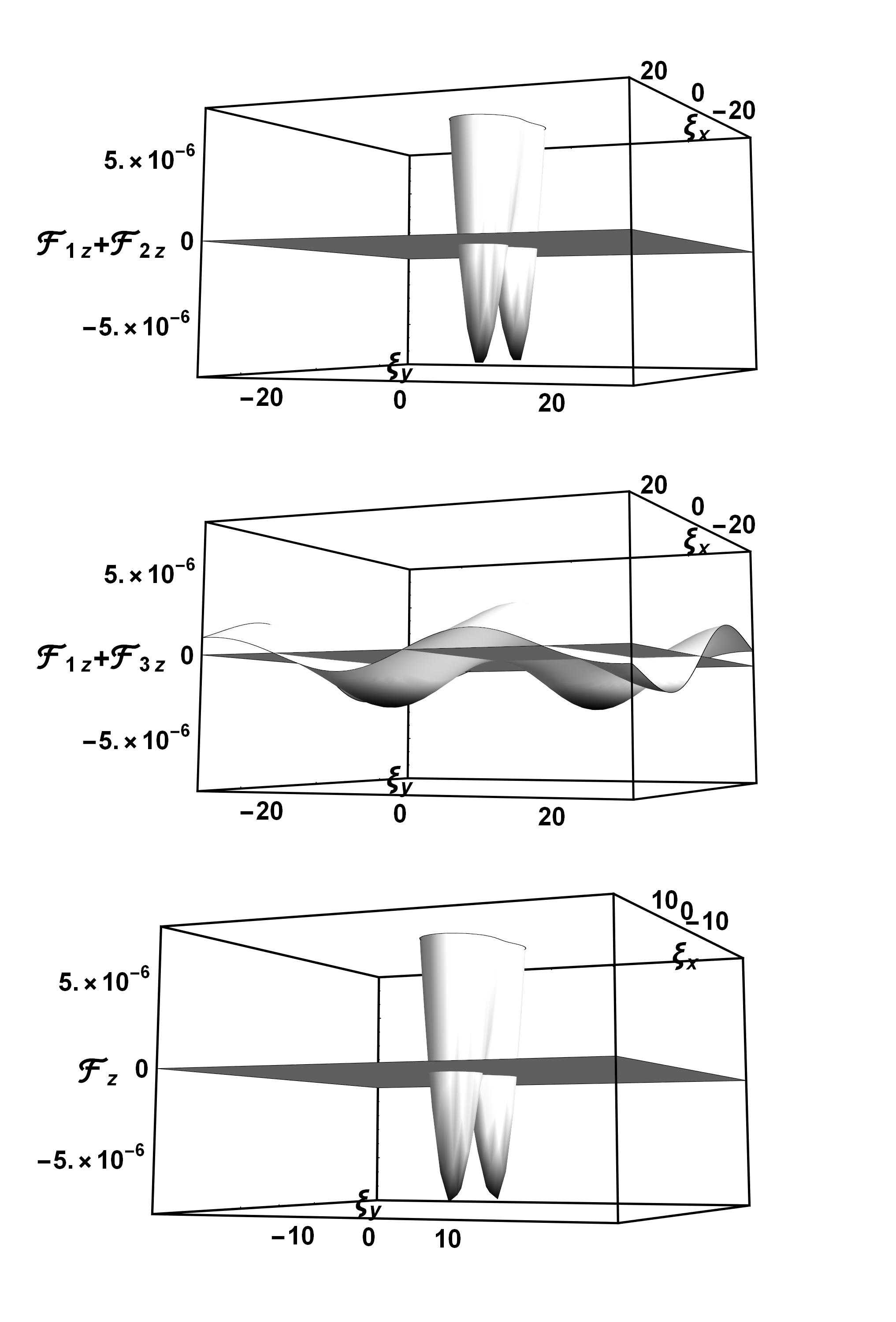}
\caption{Three-dimensional plots of selected combinations of the axial force components in the focal plane. From top to bottom: the sum of the gradient and scattering forces, the sum of the gradient and spin forces, and the total force including all three contributions. The plane at $\mathcal{F}_{z} = 0$ is highlighted to clearly indicate regions where the axial components are negative, providing insight into the spatial distribution of backward-directed forces. Parameter values are the same as in Fig.~\ref{forces2d}, except for $\beta$, which is set to $0.6$ in the present case.}
\label{allf}
\end{figure}

Using Eqs.~(\ref{resca}), (\ref{polebe}), and (\ref{parep}), the equation of motion can be written in a purely dimensionless form:
\begin{equation}\label{sical}
\ddot{\bm{\xi}}=\bm{\mathcal{F}}_{\text{grad}} + \bm{\mathcal{F}}_{\text{scatt}} + \bm{\mathcal{F}}_{\text{spin}},
\end{equation}
where the dot stands for the derivative over the dimensionless time $\tau$, i.e., $\cdot{} \equiv \partial/\partial \tau$. 
The dimensionless force components are
\begin{subequations}\label{rosic}
\begin{align}
&\bm{\mathcal{F}}_{\text{grad}} = \gamma\,\bm{\nabla} \left(\bm{\mathcal{E}}\cdot\bm{\mathcal{E}}^*\right), \label{rosicgr}\\
&\bm{\mathcal{F}}_{\text{scatt}} = 4\,\beta\,\gamma\,\bm{\mathcal{S}}, \label{rosics}\\
&\bm{\mathcal{F}}_{\text{spin}} = -\beta\,\gamma\,\bm{\nabla} \times \left(\bm{\mathcal{E}}^* \times \bm{\mathcal{E}}\right), \label{rosicsp}
\end{align}
\end{subequations}
where $\bm{\mathcal{S}}=\Re(\bm{\mathcal{E}}\times\bm{\mathcal{B}}^{*})$ is the dimensionless Poynting vector --- the factor $1/2$ arising from time averaging, as well as all dimensional constants, have been absorbed into the parameter $\gamma$ defined below in (\ref{gamma}) --- and $\bm{\nabla}$ denotes differentiation with respect to the variables $\bm{\xi}$. In order to save space in plots~\ref{forces2d} and \ref{allf}, the forces are labeled $\bm{\mathcal{F}}_1$, $\bm{\mathcal{F}}_2$, and $\bm{\mathcal{F}}_3$, corresponding to the gradient, scattering, and spin-curl components, respectively.  

Among these, the nonconservative components relevant for the present study, $\bm{\mathcal{F}}_{\mathrm{scatt}}$ and $\bm{\mathcal{F}}_{\mathrm{spin}}$, originate from the momentum flow and polarization structure of the optical field. 
In particular, the spin-curl term reflects spin–orbit coupling, whereby polarization gradients generate local forces even in regions of nearly uniform intensity.

Within the present paraxial model, backward-directed optical forces on Rayleigh particles can emerge through two distinct mechanisms:
\begin{enumerate}
\item \textbf{Poynting-vector backflow} -- local reversals of the energy flux $\bm{\mathcal{S}}$, acting through the scattering term $\bm{\mathcal{F}}_{\mathrm{scatt}}$.  
\item \textbf{Spin–momentum coupling} -- polarization-de\-pen\-dent curl force associated with spatial variations of the spin density within $\bm{\mathcal{F}}_{\mathrm{spin}}$, capable of driving the particle backward even when the net energy flow remains forward.
\end{enumerate}

Both mechanisms vanish in the lowest-order scalar paraxial approximation and appear only when higher-order vector corrections (of order $\varepsilon^2$ or higher) are included to satisfy the transversality conditions more accurately~\cite{trmax}. Backward optical forces are then genuine manifestations of the vector nature of light and of the coupling between its canonical and spin momentum densities~\cite{beksa2,beksa,shi}.

All three forces are proportional to the same small parameter $\gamma$,
\begin{equation}\label{gamma}
\gamma \equiv \frac{\alpha \mathcal{E}_0^2}{4 (1+\beta^2) m c^2},
\end{equation}
so its exact value is not essential for comparison purposes.  
Therefore, in the plots of this section, which are intended to illustrate the relative magnitudes of the three forces, one can simply set $\gamma = 1$.  

\begin{figure*}[t!]
\begin{center}
\includegraphics[width=0.95\textwidth,angle=0]{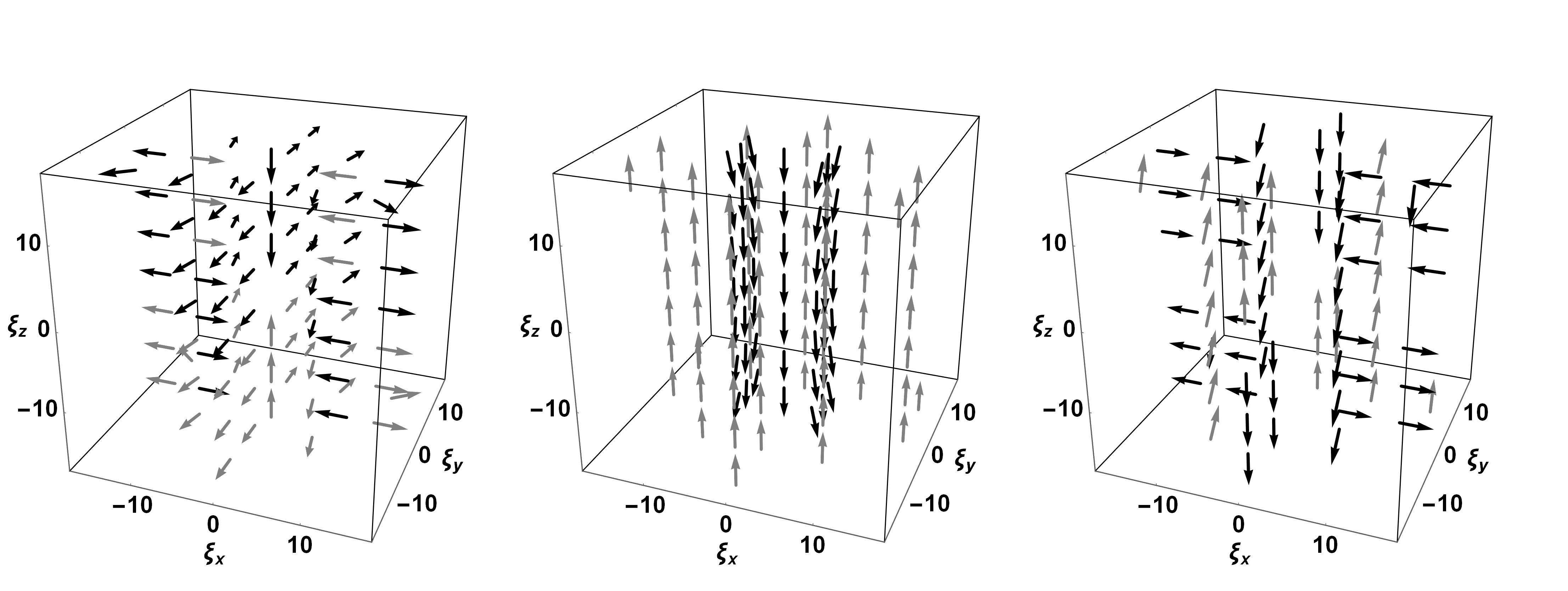}
\caption{Spatial distributions of the gradient (left), scattering (middle), and spin (right) forces, represented as vector fields. Black arrows indicate regions where the axial component of the force is negative (even if small), whereas gray arrows correspond to positive axial components. The parameter values are the same as in Fig.~\ref{forces2d}. Arrow lengths are not scaled to the force magnitudes.}
\label{forces23}
\end{center}
\end{figure*}

The situation changes when the particle’s motion is taken into account, since under realistic conditions the parameter $\gamma$ becomes extremely small.  
For a Rayleigh particle of radius $R \approx 100\, \text{nm}$, refractive index $n \approx 1.6$, and density $\rho \approx 10^3\, \text{kg/m}^3$, illuminated by a laser light of intensity $I \approx 10^{10}\,\text{W/m}^2$ and wavelength $\lambda = 532\, \text{nm}$, one obtains $\gamma \approx 1.7\times10^{-19}$ and $\beta \approx 0.38$.  
At this level, optical forces are dominated by thermal noise, and the particle motion is strongly influenced by Brownian fluctuations. Deterministic effects associated with the optical forces can then be revealed only as systematic trends superimposed on stochastic motion, and become experimentally observable primarily under cryogenic temperatures or in high vacuum, where viscous damping and Brownian noise are strongly suppressed.  
In such conditions, again only the relative balance between the forces is meaningful.

For numerical visualization of the particle’s trajectories (in the following section), we therefore decided to set $\gamma = 10^{-4}$, which -- although still much larger than the physical value -- preserves the correct relative strengths of the forces while providing a convenient dynamical scale and avoiding excessively long integration times for very small $\gamma$. At this value, the interplay between the gradient, scattering, and spin–curl forces becomes clearly visible in the simulated trajectories. Larger effective values of $\gamma$ could, in principle, be achieved by stronger illumination, tighter focusing, or by using materials with a higher refractive index.

Let us first examine the gradient force in the vicinity of the propagation axis, focusing in particular on its $\xi_z$ component.
By substituting Eq.~(\ref{parpee}) into Eq.~(\ref{rosicgr}), using Eq.~(\ref{vv}), and retaining only the lowest-order terms in $\varepsilon$ (except for exponential factor), one obtains for $\delta=1$ (the full expression is too lengthy to be reproduced here)
\begin{eqnarray}\label{gradfr}
\mathcal{F}_{\text{grad},z} \approx&& -96\gamma\varepsilon^8\xi_z\,e^{-\frac{2\varepsilon^2\xi_\perp^2}{1+4\varepsilon^4\xi_z^2}}\Big\{9+10\varepsilon(1-\xi_x^2+\xi_y^2)\nonumber\\
&&+\varepsilon^2\left[1+\xi_\perp^4-4\,(11\xi_x^2-36\xi_y^2)\right]\Big\}.
\end{eqnarray}
Here, $\xi = |\boldsymbol{\xi}|$ and $\xi_\perp = \sqrt{\xi_x^2+\xi_y^2}$.
It is evident that this component changes sign below and above the plane $\xi_z = 0$, which is a natural and expected behavior.
The focal plane ($\xi_z = 0$) corresponds to the region of maximal energy concentration. Consequently, a particle located either above or below this plane experiences a restoring gradient force that drives it back toward it.
This behavior demonstrates the trapping nature of the optical field along the propagation direction and identifies the focal plane as a point of equilibrium.
Naturally, the actual three-dimensional optical trap is not formed on the axis itself, but rather on a ring surrounding it, as shown in Fig.~\ref{inty}.

It should be pointed out that if the electric and magnetic fields did not include the additional terms identified in Ref.~\cite{trmax} -- that is, for $\delta = 0$ -- the axial component of the gradient force would vanish near the focus. In this case, the exact expression for $\mathcal{F}_{\text{grad},z}$ reads
\begin{eqnarray}\label{gradfr0}
\mathcal{F}_{\text{grad},z}\Big|_{\delta=0} = &&-32\gamma\varepsilon^{10}\xi_z\xi_\perp^4 e^{-\frac{2\varepsilon^2\xi_\perp^2}{1 + 4\varepsilon^4\xi_z^2}}\\
&&\times\frac{(3 - 2\varepsilon\xi_\perp^2 + 12\varepsilon^4\xi_z^2)}{(1 + 4\varepsilon^4\xi_z^2)^5}\,.\nonumber
\end{eqnarray}
Thus, in the absence of higher-order field corrections, the restoring force along $\xi_z$ appears only at the fourth order in the transverse displacement.  
In the physically relevant case $\delta = 1$, this force remains nonzero but is strongly suppressed by a factor of $\varepsilon^8$ as seen in (\ref{gradfr}).  
As will be shown below, in this same region a reversal of the particle’s motion may occur, driven by the backflow of the Poynting vector.  
Further away from the axis, however, the backward motion of the particle may also be assisted by the spin-dependent component of the optical force.

The scattering force along the $\xi_z$ direction, upon setting $\delta=1$, is given by
an expression [derived using the same approximation as in~(\ref{gradfr})]
\begin{eqnarray}\label{sfr}
\mathcal{F}_{\text{scatt},z}
&\approx& -4\beta\gamma\varepsilon^5e^{-\frac{2\varepsilon^2\xi_\perp^2}{1+4\varepsilon^4\xi_z^2}}\Big\{6(2+\xi_x^2-\xi_y^2)\nonumber\\
&&+2\varepsilon(2-\xi_x^2+\xi_y^2-\xi_\perp^4)\\
&&-3\varepsilon^2\left[14 \xi_x^2 + 34 \xi_y^2 + (3\xi_x^2+13 \xi_y^2)(\xi_x^2 - 
 \xi_y^2)\right]\Big\}.\nonumber
\end{eqnarray}

This component is negative on the axis, both above and below the focal plane, and scales as $\varepsilon^{5}$, which makes it relatively large compared with the corresponding axial gradient force in the same area. 
Its sign indicates that the scattering force tends to push the particle backward, opposite to the nominal propagation direction of the beam.  
Importantly, this effect arises entirely from the additional field terms that modify the local structure of the Poynting vector.  
In the absence of these corrections (i.e., for $\delta = 0$) the characteristic backflow of the Poynting vector responsible for this effect no longer occurs.  
In that case, the full expression for $\mathcal{F}_{\text{scatt},z}$ simplifies to
\begin{equation}\label{sfr0}
\mathcal{F}_{\text{scatt},z} = \frac{8\beta\gamma\varepsilon^6\xi_\perp^4}{(1 + 4\varepsilon^4\xi_z^2)^3}\,
e^{-\frac{2\varepsilon\xi_\perp^2}{1 + 4\varepsilon^4\xi_z^2}},
\end{equation}
and remains strictly positive. Therefore, without the higher-order paraxial field terms, the scattering force acts only in the forward direction, precluding any backward-directed motion.

The $\xi_z$ component of the scattering force can be also viewed by considering the total flux of the force through a transverse plane at fixed $\xi_z$.  The integrated force is independent of $\xi_z$ and given by
\begin{eqnarray}\label{casi}
\iint\! \mathcal{F}_{\text{scatt},z}&&\, d^2\xi_\perp = \\
&&\pi\beta\gamma\left[2 - \frac{3}{4}\,\delta^2\varepsilon(1-\varepsilon) - 3\delta^4\varepsilon^3(2+\varepsilon)\right].\nonumber
\end{eqnarray}
Naturally, this determines only the overall sign of the flux, which remains positive and corresponds to forward-directed momentum transfer.  
However, this does not preclude the existence of localized regions where the force acts in the opposite direction. Notably, the presence of the two negative terms for $\delta \neq 0$ reflects these local backward effects, revealing the subtle spatial structure of the scattering force even within a nominally forward-propagating beam.

The expression for the $\xi_z$ component of the spin force for small $\varepsilon$ and $\delta=1$ is given by
\begin{equation}\label{spfr}
\mathcal{F}_{\text{spin},z} = 
128\beta\gamma\varepsilon^6\xi_x\xi_y
e^{-\frac{2\varepsilon\xi_\perp^2}{1+4\varepsilon^4\xi_z^2}}
\left[1-\frac{\varepsilon^2}{2}(\xi_x^2+6\xi_y^2)\right].
\end{equation}

In the central part of the beam, this component does not contribute to the backward motion.  
It should also be emphasized that the spin force vanishes completely when $\delta = 0$, confirming that it originates solely from the higher-order field corrections.  
Nevertheless, at certain off-axis positions where the product $\xi_x\xi_y<0$ or when the second term dominates the first, the spin force becomes negative and thus assists the reversed motion of the particle. 

\begin{figure*}
\begin{center}
\includegraphics[width=0.95\textwidth]{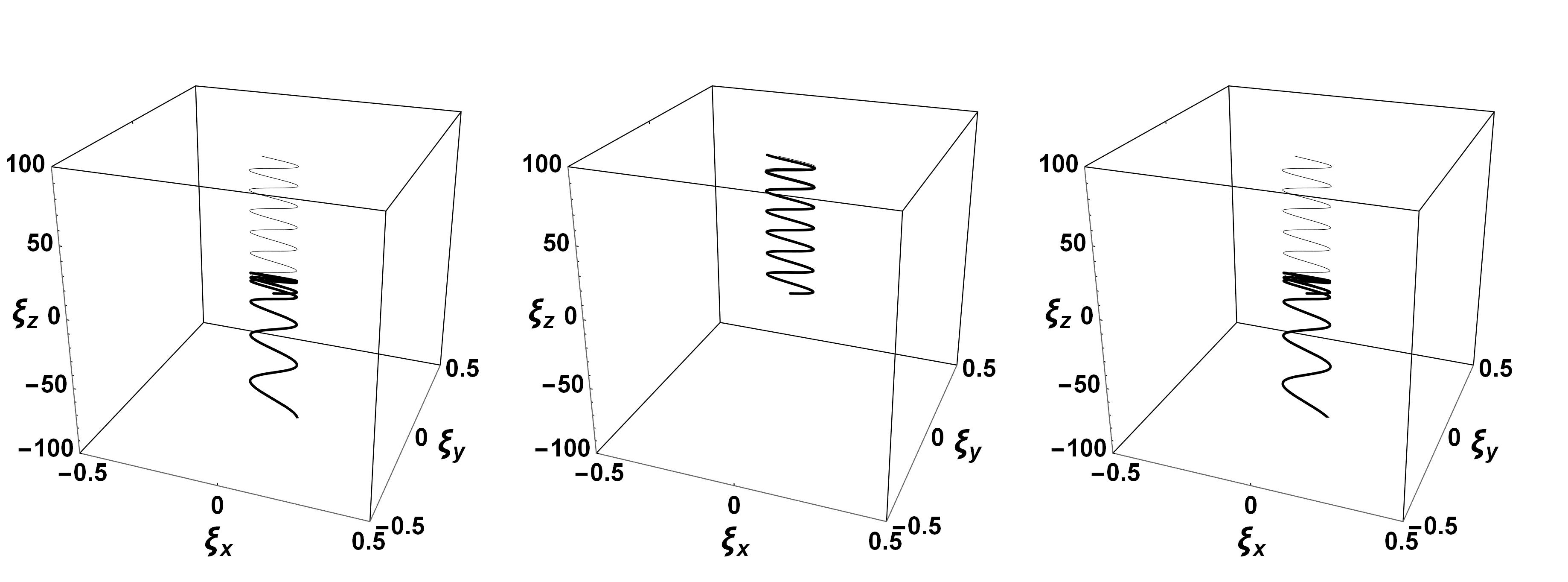}
\caption{Representative trajectories of a Rayleigh particle obtained from numerical integration of Eq.~(\ref{sical}) for different combinations of optical forces. From left to right: (i) the gradient and scattering forces together (thick line) compared with the gradient force alone (thin line); (ii) the gradient and spin forces, for which the two trajectories almost coincide; and (iii) the total force including all three components (thick line) compared with the gradient force (thin line). In the first and third cases, a clear reversal of motion appears, showing backward pulling caused by the negative axial component of the Poynting vector. The simulation parameters are $\varepsilon = 0.03$, $\gamma = 10^{-4}$, and $\beta = 0.3$.}
\label{trajscatt}
\end{center}
\end{figure*} 

To illustrate this behavior, the full expression for $\mathcal{F}_{\text{spin},z}$ in the focal plane under the simplifying condition $\xi_y = -\xi_x$ is given below:
\begin{equation}\label{spfr00}
\mathcal{F}_{\text{spin},z}
= -16\beta\gamma\delta\varepsilon^6\xi_x^2
e^{-4\varepsilon\xi_x^2}
\left[\left(4\varepsilon^2\xi_x^2-7\right)^2-41\right].
\end{equation}
This expression shows that the force becomes negative for 
$0 < |\xi_x| \lesssim \frac{0.39}{\varepsilon}$, 
indicating that the spin contribution indeed drives the particle backward within this range of transverse displacements.
The total flux of the spin-curl force through any transverse plane at fixed $\xi_z$ vanishes identically. This result follows from the mathematical structure of this term, which can be written as a curl of the local spin density. The spin-curl force redistributes momentum in the transverse directions without contributing to the overall forward-directed flux along the propagation axis. Consequently, while local backward motion can appear in off-axis regions due to the spin-curl term, the net momentum along the beam remains unchanged.

Fig.~\ref{forces2d} compares the axial components of all three forces in the near-axis region, showing their relative strengths and spatial variations.  
The numerical results are fully consistent with the analytical expressions discussed above and clearly demonstrate the possibility of backward motion caused by the Poynting and spin-related terms.  
The lower panel presents the forces calculated from the full expressions (\ref{parpee}) and (\ref{parpbe}), while the upper panel corresponds to the case $\delta = 0$.  
In this case -- which is mainly formal and physically unrealistic, since the paraxial fields are not accurate enough -- all backward-directed forces are absent. 

For the parameters considered, the backflow region has a diameter of approximately $6/k \approx \lambda$. Thus, for a Rayleigh particle with a characteristic size of several tens of nanometers, or roughly $100\, \mathrm{nm}$ at a wavelength of $1064\, \mathrm{nm}$, the dipole approximation remains valid. The extent of this region can be further tuned by adjusting the value of $\varepsilon$, which, however, also affects the absolute magnitude of the forces.

In turn, Fig.~\ref{forces23} provides a comprehensive visualization of the scattering and spin forces, highlighting the local directions of the corresponding vectors. From the arrow patterns, one can easily identify the regions where they act opposite to the beam propagation, thus illustrating the spatial structure of the backward-directed interaction and its possible role in reversing the particle’s motion.

Finally, Fig.~\ref{allf} displays three-dimensional plots of different combinations of the axial force components: the sum of the gradient and scattering forces, the sum of the gradient and spin forces, and the total force including all three contributions.  
All plots are shown in the focal plane, with the level corresponding to $\mathcal{F}_{z} = 0$ marked to clearly indicate the regions where the axial components of the forces are negative.

\begin{figure}[b]
\begin{center}
\includegraphics[width=0.35\textwidth,angle=0]{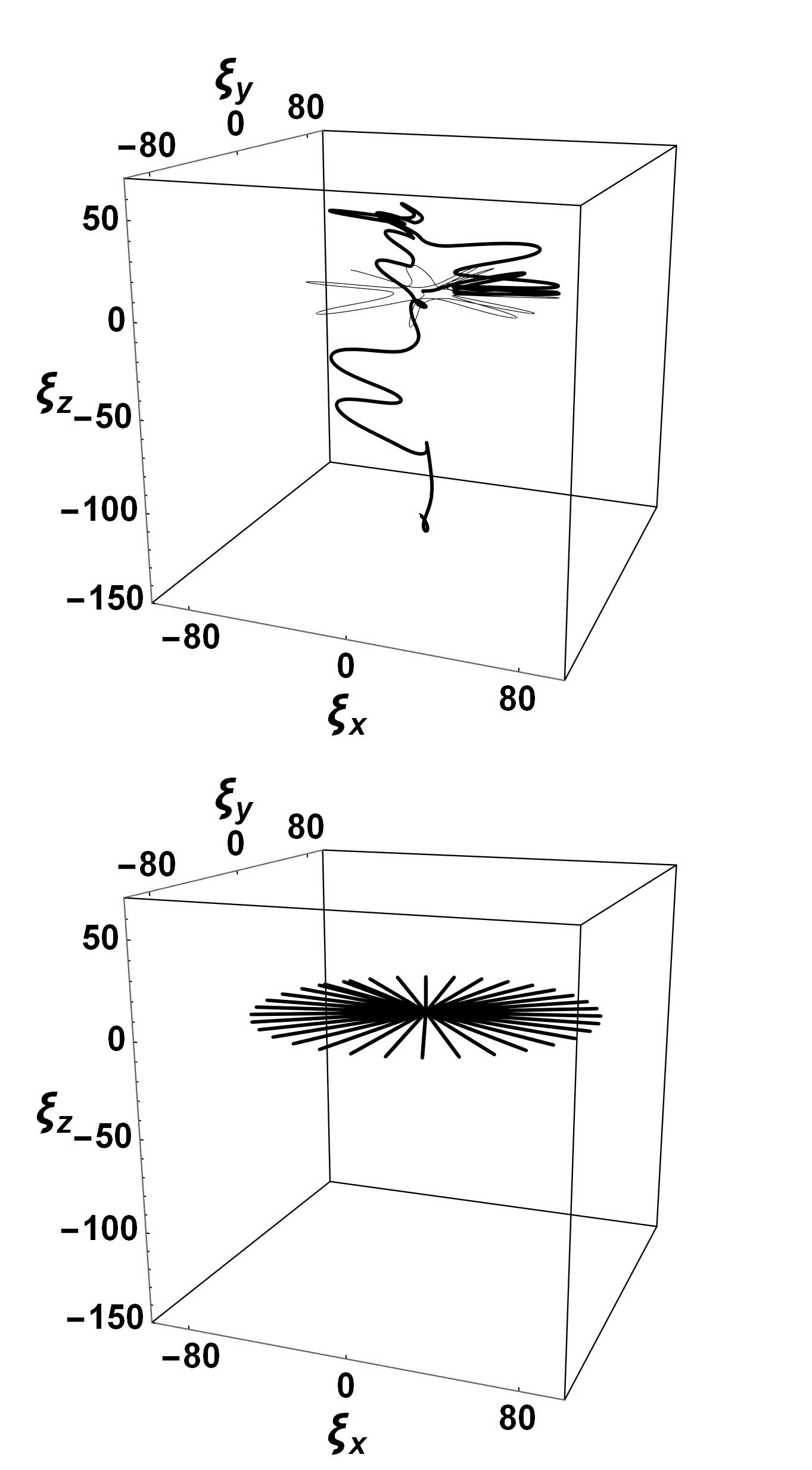}
\caption{Representative trajectories of a Rayleigh particle obtained by numerical integration of Eq.~(\ref{sical}) in the presence of the spin-related force.  
Upper panel: comparison between the combined gradient–spin dynamics (thick line) and the motion driven solely by the gradient force (thin line).  
The particle is initially placed off the optical axis, in a region where both the gradient and spin components are significant ($\xi_x=-2.5$, $\xi_y=2.5$ and $\xi_z=0$). Parameter values are $\varepsilon = 0.1$, $\gamma = 10^{-4}$, and $\beta = 0.1$. The spin force introduces noticeable transverse perturbations, occasionally leading to irregular or partially chaotic (in general backward) motion.  
Lower panel: the same configuration for $\delta = 0$, where the higher-order paraxial corrections vanish and the trajectory does not exhibit any reversal.}
\label{trajsp}
\vspace{-1ex}
\end{center}
\end{figure}

\section{Numerical simulations of particle motion}\label{numtr}

Having drawn the spatial maps of $\mathcal{F}_{\text{scatt},z}$ and $\mathcal{F}_{\text{spin},z}$, which revealed extended regions where these components take negative values, we now turn to the numerical analysis of particle trajectories to verify whether they indeed exhibit backward pulling. The trajectories shown below were obtained by numerically integrating the equation of motion~(\ref{sical}).

Figure~\ref{trajscatt} presents three representative examples illustrating the motion of a Rayleigh particle under different combinations of optical forces. In the first case, corresponding to the combined action of the gradient and scattering forces, a clear reversal of motion is observed relative to the trajectory driven solely by the gradient term. The particle initially moves forward under the influence of the gradient force, since its starting position is chosen slightly below the focal plane and the initial velocity is directed upward. Subsequently, the presence of a negative component of the Poynting vector causes it to reverse direction and move backward.

In the second configuration, where the gradient and spin forces act together while the scattering contribution is switched off, the particle follows a smooth forward path without any reversal, indicating that the spin term is negligible in this region. This conclusion is consistent with Eq.~(\ref{spfr}) and with the spatial distributions shown in Figs.~\ref{forces2d}, \ref{forces23}, and~\ref{allf}. The transverse motion remains confined near the optical axis, preventing the particle from entering regions where $\mathcal{F}_{\text{spin},z}$ attains appreciable values.

When all three forces are included, the resulting trajectory closely follows that obtained for the gradient–scattering combination, confirming that in this regime the backward motion is mainly driven by the scattering force. It should also be noted -- which is not shown in the figures, but verified numerically -- that for $\delta = 0$ the particle quickly escapes the interaction region, with neither trapping nor backward pulling occurring.

In the simulations discussed above, the dimensionless parameter $\beta$ was set to $0.3$. 
This value corresponds to a moderate imaginary part of the particle’s complex polarizability and reflects a balance between conservative (gradient) and dissipative (scattering) optical forces. 
Physically, $\beta$ is proportional to the ratio of the imaginary to the real part of the polarizability, $\beta \sim \Im(\alpha)/\Re(\alpha)$, and therefore depends on the material composition, wavelength, and particle size. For weakly absorbing dielectric particles commonly used in optical trapping experiments, such as doped silica or low-loss polymer beads, typical values of $\beta$ are significantly smaller than unity but non-negligible. In contrast, metallic or resonant particles can exhibit much larger imaginary polarizability associated with plasmonic absorption and scattering. Examples of the real and imaginary parts of the complex dipole polarizability for metallic (gold) nanoparticles, from which representative values of $\Im(\alpha)/\Re(\alpha)$ can be inferred, are provided in~\cite{baby}.

Figure~\ref{trajsp} further illustrates the dynamics in the presence of the spin-related force. The upper panel compares two trajectories: one obtained under the combined action of the gradient and spin forces (thick line) and another corresponding to the pure gradient force (thin line). Here, the particle is initially placed farther from the optical axis, in a region where both forces exhibit larger amplitudes. The parameter $\beta$ is set to $0.1$, corresponding to a weakly dissipative particle dominated by conservative optical effects.

As seen in the plot, the spin force introduces substantial transverse perturbations, leading to less regular, partially chaotic motion that may even expel the particle from the focal region. Indeed, in this regime, defining a stable trajectory when all three forces are included proves challenging, as the chaotic character induced by the spin term can quickly push the particle into the region where the scattering force dominates. Nevertheless, compared with the smooth confinement provided by the gradient force alone (thin line), the presence of the spin term generates a subtle backward drift superimposed on the otherwise forward propagation. This behavior reflects the influence of local spin–orbit coupling effects, which alter the topology of the Poynting vector and give rise to small-scale optical backflow.

The lower panel shows the corresponding simulation performed for $\delta = 0$. In this case, the beam structure becomes effectively symmetric, resulting in regular motion without any reversal or noticeable spin-induced perturbations. The comparison of both panels demonstrates that the observed backward and irregular trajectories arise solely from higher-order, polarization-dependent corrections encoded in the spin force.

\section{Conclusions}

We have presented a combined theoretical and numerical study of optical forces on a Rayleigh particle in a paraxial Gaussian beam exhibiting optical backflow. Within the dipole approximation, the total force is decomposed into gradient, scattering, and spin-curl components, with the latter two responsible for backward-directed motion.

Using vector fields that satisfy the exact paraxial Maxwell equations, we show that higher-order terms are essential: neglecting them leaves the Poynting vector strictly forward, eliminating backward contributions. Including these terms produces regions of negative longitudinal energy flux, reversing the scattering force near the beam axis, while the spin-curl term generates weaker, off-axis backward contributions linked to local spin density variations.

The strength of these effects depends on the dimensionless parameters $\gamma$ and $\beta$, which scale the optical forces and the nonconservative contributions. Backward components vanish in the absence of dissipation.

Numerical simulations confirm that near-axis backward motion is dominated by the scattering force, whereas the spin-curl term induces localized off-axis deviations. The combination explains the complex, sometimes partially reversed particle trajectories observed in structured paraxial fields.

This framework offers a basis for extending studies to nonparaxial beams, larger particles, or engineered spin–orbit interactions, and for exploring the interplay of backflow-mediated forces with stochastic effects such as Brownian motion, with potential applications in thermally assisted particle manipulation.

\end{document}